\documentclass[acmsmall]{acmart}
\usepackage{multirow}
\usepackage{color, colortbl, xcolor}

\newcommand{\rowcollight}{\rowcolor{LightGray}}
\definecolor{LightGray}{gray}{0.97}

\usepackage{longtable}
\usepackage{capt-of}
\usepackage{balance,bm}
\usepackage{color, colortbl, xcolor}
\definecolor{LightGray}{gray}{0.97}
\usepackage{url}
\usepackage{hyperref}
\hypersetup{
    colorlinks=true,
}

\usepackage{subcaption}
\usepackage{booktabs} % For formal tables
\usepackage{graphicx}
\usepackage{textcomp}
\usepackage{color,soul}
\usepackage{bm}
\usepackage{multirow}
\usepackage{balance}
\usepackage{graphicx}  %Required
\usepackage{enumitem}

\usepackage{colortbl}
\usepackage{arydshln}
\usepackage [english]{babel}
\usepackage [autostyle, english = american]{csquotes}
\definecolor{linkColor}{RGB}{6,125,233}
\definecolor{green}{rgb}{0.0, 0.65, 0.31}
\definecolor{bleudefrance}{rgb}{0.19, 0.55, 0.91}
\definecolor{ceruleanblue}{rgb}{0.16, 0.32, 0.75}
\definecolor{grey}{HTML}{969696}
\definecolor{violet}{HTML}{756bb1}
\definecolor{dgrey}{HTML}{01665e}
\definecolor{lgrey}{HTML}{5ab4ac}
\definecolor{dgreen}{HTML}{005a32}
\definecolor{purple}{HTML}{ae017e}

 \definecolor{editCol}{HTML}{000000}
\definecolor{maskCol}{HTML}{c51b7d}
\definecolor{lrColor}{HTML}{8856a7}
\definecolor{trColor}{HTML}{d01c8b}
\definecolor{ctColor}{HTML}{4dac26}
\definecolor{brickred}{HTML}{f03b20}
\definecolor{improveCol}{HTML}{253494}
\definecolor{worsenCol}{HTML}{d7191c}
\definecolor{DarkBlue}{HTML}{00008B}
\definecolor{mscolor}{HTML}{01665e}
\definecolor{nmscolor}{HTML}{bf812d}
\definecolor{lgreen}{HTML}{ccece6}
\definecolor{dolive}{HTML}{308014}

\definecolor{maskCol}{HTML}{c51b7d}
\definecolor{lrColor}{HTML}{8856a7}
\definecolor{trColor}{HTML}{d01c8b}
\definecolor{ctColor}{HTML}{4dac26}
\definecolor{brickred}{HTML}{f03b20}
\definecolor{improveCol}{HTML}{253494}
\definecolor{worsenCol}{HTML}{d7191c}
\definecolor{lgreen}{HTML}{e0f3db}
\definecolor{dpink}{HTML}{CD1076}
\definecolor{pink}{HTML}{FED2D2}
\definecolor{soothinggreen}{HTML}{4dac26}
\definecolor{darkred}{HTML}{8B0000}

\definecolor{dblue}{HTML}{104E8B}
\definecolor{violet}{HTML}{8A2BE2}
\definecolor{mscolor}{HTML}{01665e}
\definecolor{nmscolor}{HTML}{d8b365}
\definecolor{deepgrey}{HTML}{525252}
\definecolor{dslate}{HTML}{2F4F4F}
\definecolor{dolive}{HTML}{556B2F}
\definecolor{teal}{HTML}{388E8E}
\definecolor{mscolor}{HTML}{01665e}
\definecolor{nmscolor}{HTML}{d8b365}

\definecolor{aicolor}{HTML}{018571}
\definecolor{occolor}{HTML}{ff7799}

\definecolor{srcolor}{HTML}{e34a33}
\definecolor{smcolor}{HTML}{253494}
\definecolor{srsmcolor}{HTML}{7fcdbb}
\definecolor{bothcolor}{HTML}{fe9929}
\definecolor{onecolor}{HTML}{018571}
\definecolor{marroon}{HTML}{881c1c}

\usepackage{mathtools}

\usepackage{amsmath}
\usepackage{array}
\usepackage{xcolor}
\usepackage{arydshln}
\usepackage{siunitx}
\colorlet{tablerowcolor4}{gray!50} % Table row separator colour = 

\newcommand*{\textlabel}[2]{%
  \edef\@currentlabel{#1}% Set target label
  \phantomsection% Correct hyper reference link
  #1\label{#2}% Print and store label
}

\colorlet{tableheadcolor}{gray!25} % Table header colour = 25% gray
\colorlet{tablerowcolor}{gray!15} % Table row separator colour = 
\colorlet{tablerowcolor2}{gray!45} % Table row separator colour = 
\colorlet{tablerowcolor3}{gray!25} % Table row separator colour = 10% gray

\newif{\ifhidecomments}
  \hidecommentsfalse 
\ifhidecomments
    \newcommand{\melissa}[1]{}
    \newcommand{\dongwhi}[1]{}
    \newcommand{\koustuv}[1]{}
    \newcommand{\violeta}[1]{}
\else
    
    \newcommand{\melissa}[1]{\textbf{\small\sffamily{\textcolor{dolive}{[#1 -- Melissa]}}}}
    \newcommand{\dongwhi}[1]{\textbf{\small\sffamily{\textcolor{dpink}{[#1 -- Dong Whi]}}}}
    \newcommand{\koustuv}[1]{\textbf{\small\sffamily{\textcolor{violet}{[#1 -- Koustuv]}}}}
  \fi
\newcommand{\edit}[1]{{\textcolor{editCol}{#1}}}

\definecolor{neutralCol}{HTML}{dd1c77}
\definecolor{neutralGreen}{HTML}{31a354}
\definecolor{NewBlue}{HTML}{1879ba}
\definecolor{bleudefrance}{rgb}{0.19, 0.55, 0.91}  
\definecolor{AfTrColor}{HTML}{0868ac}  
\definecolor{BfTrColor}{HTML}{a8ddb5}  

\definecolor{AfCtColor}{HTML}{b10026}  
\definecolor{BfCtColor}{HTML}{fd8d3c}

\graphicspath{ {figures/} }

\renewcommand{\textcolor}[2]{#2}

\AtBeginDocument{%
  }

\setcopyright{cc}
\setcctype{by}
\acmJournal{PACMHCI}
\acmYear{2026} \acmVolume{10} \acmNumber{6} \acmArticle{CSCW164}
\acmMonth{10} \acmDOI{10.1145/3817012}

\begin{document}

\title[Tensions between the FDA’s Data Governance Vision and the Lived Realities of Food Producers]{``The New Era of Tech-Enabled Traceability'': Tensions between the FDA’s Data Governance Vision and the Lived Realities of Food Producers}

\author{Soonho Kwon}
\orcid{0000-0002-2783-6364}
\affiliation{%
  \institution{Georgia Institute of Technology, School of Interactive Computing}
  \city{Atlanta, GA}
  \country{USA}
  \postcode{30332}
  }
  \email{soonho@gatech.edu}

\author{Catherine Wieczorek}
\orcid{0000-0002-1251-7694}
\affiliation{%
  \institution{Georgia Institute of Technology, School of Interactive Computing}
  \city{Atlanta, GA}
  \country{USA}
  \postcode{30332}
  }
  \email{cwieczor3@gatech.edu}

\author{Heidi Biggs}
\orcid{0000-0001-9208-7921}
\affiliation{%
  \institution{Georgia Institute of Technology, School of Literature Media and Communication}
  \city{Atlanta, GA}
  \country{USA}
  \postcode{30332}
  }
  \email{hbiggs7@gatech.edu}

\author{Shellye Suttles}
\orcid{0000-0003-4100-8431}
\affiliation{%
  \institution{Indiana University Bloomington, O’Neill School of Public \& Environmental Affairs}
  \city{Bloomington, IN}
  \country{USA}
  \postcode{47405}
  }
  \email{shelsutt@iu.edu}

\author{Tammi S. Etheridge}
\orcid{0009-0001-0902-7045}
\affiliation{%
  \institution{Washington and Lee University School of Law}
  \city{Lexington, VA}
  \country{USA}
  \postcode{24450}
  }
  \email{tetheridge@wlu.edu}

\author{Annabel Rothschild}
\orcid{0000-0002-5882-1608}
\affiliation{%
  \institution{Bard College, Division of Sciences, Mathematics, and Computing}
  \city{Annandale-on-Hudson, NY}
  \country{USA}
  \postcode{12504}
  }
  \email{arothschild@bard.edu}

\author{Shaowen Bardzell}
\orcid{0000-0001-7596-9244}
\affiliation{%
  \institution{Georgia Institute of Technology, School of Interactive Computing}
  \city{Atlanta, GA}
  \country{USA}
  \postcode{30332}
  }
  \email{shaowen@cc.gatech.edu}

\renewcommand{\shortauthors}{Kwon et al.}

\begin{abstract}
The U.S. Food and Drug Administration (FDA)’s Food Traceability Rule, promulgated as part of the Food Safety Modernization Act (FSMA), requires \textit{agri-food supply chain stakeholders (stakeholders)}—including farmers, fishers, retail workers, and others—to maintain detailed tracking records beginning in January 2026. Through this Rule, the FDA envisions a “New Era of Tech-Enabled Traceability,” in which standardized, harmonized tracking data serve as a foundational public health infrastructure, enabling more rapid identification and removal of potentially contaminated food and ultimately reducing the risk of foodborne illness.

Despite this promising vision, we observe that the Rule reconfigures agri-food stakeholders into data laborers by mandating stringent data-collection, formatting, and reporting requirements. In this paper, we examine the tensions and burdens that arise from such reconfiguration. Leveraging Data Feminism as an orientation to attend to how data-driven policy implementation disproportionately burdens smaller, under-resourced stakeholders who lack the infrastructural and financial capacity to comply, we analyze 1,198 public comments submitted to Regulations.gov in response to the proposed Rule. Our qualitative document analysis reveals three key tensions: (1) the individual labor, financial, and educational burdens stakeholders experience as they are reconfigured into data workers; (2) moments where data tracking becomes infeasible due to infrastructural limitations, cultural contexts, and situated production practices; and (3) instances where the Rule’s intended flexibility instead introduces confusion and burden due to its ambiguity.

This work contributes to CSCW by providing an empirical account of how data-driven policy reconfigures stakeholders as data producers. We demonstrate that those reconfigured as data workers under data-driven governance must be recognized as meaningful \emph{policy partners} rather than merely as actors subjected to compliance obligations. We further suggest that CSCW can play a critical role in uncovering their yet invisible burdens, supporting their capacity to voice concerns through participatory democratic mechanisms, and aiding their ability to “make-do” under imposed data works.
\end{abstract}




\maketitle

\section{Introduction}

In response to growing concerns about foodborne illness in the United States, the Food and Drug Administration (FDA) implemented the Food Safety Modernization Act (FSMA) in 2011, under the vision of creating a “new era of tech-enabled traceability.” FSMA marked a shift from reactive to preventive food safety, motivated by evidence that millions of preventable illnesses and hundreds of deaths annually were linked to contaminated food. Among its nine major rules, the FSMA Final Rule on Requirements for Additional Traceability Records for Certain Foods (the Food Traceability Rule) mandates that stakeholders handling certain fresh foods maintain detailed, standardized data related to their food production practices by early 2026, in order to enable faster identification and removal of contaminated products \cite{fda2022traceability}.

In this paper, we argue that this data-driven governing vision and its implementation in food production reconfigure stakeholders into data workers. We use the term \textit{agri-food supply chain stakeholders (stakeholders)} to denote the complexity of the food system, which includes farmers, fishers, distributors, food processors, retailers, and regulatory actors operating at different scales. We illustrate how they are transformed into \emph{data workers} tasked with maintaining granular compliance with complex informational infrastructures alongside their existing professional responsibilities. We further demonstrate how this reconfiguration creates significant tensions and burdens for stakeholders as they seek to comply with the newly introduced data work requirements imposed by the Food Traceability Rule.

Studies in CSCW have shown that data practices across diverse sectors, such as corporations, hospitals, and schools, often rely on hidden, unacknowledged forms of labor distributed unevenly among stakeholders with differing power and resources \cite{passi2017data,lu2021data,muller2019data}. Attending to these dynamics not only reveals \textit{who} bears the burden of data-driven infrastructures but also foregrounds the \textit{tensions} between abstract visions of efficiency, the implementation of such visions through data-driven practices, and the lived realities of those who must perform the data work \cite{sum2025s,bopp2017disempowered}. In food production \edit{specifically}, \citeauthor{doggett2024migrant} give special attention to how policy infrastructures and legal platforms shape farmers' entanglements with data-driven agriculture, discussing how farmers' data practices are constrained and reshaped by policy mandates and infrastructural obligations that require their participation in such data-driven agricultural practices \cite{doggett2024migrant}.

In this paper, we build on these efforts by examining specific tensions between the Food Traceability Rule's vision, its implementation practices, and the lived realities of stakeholders who must comply, as well as the new burdens that arise from these tensions. To do so, we conducted a qualitative document analysis of 1,198 public comments submitted via Regulations.gov by stakeholders. On this digital platform, federal agencies provide details on proposed and finalized regulations and allow the public (both individuals and organizations) to engage in the rulemaking process by submitting comments on proposed rules. We orient our analysis based on the first principle of Data Feminism, ``examine power'' \cite{datafeminism} by treating these public comments not as passive feedback but as active expressions of concern, critique, and resistance to top-down decision-making rooted in stakeholders' lived realities.

Our findings highlight three specific tensions that arise when stakeholders are reconfigured into data workers through the FDA's data-driven policy vision: (1) at the individual level, stakeholders face significant educational, labor, and financial burdens as they reshape their identities and work practices to fit data work rather than their original task of food production; (2) at the production environment level, stakeholders encounter situations in which performing the data work required by the policy is impossible or infeasible due to physical or cultural infrastructures, as well as on-the-ground practices; and (3) at the policy level, stakeholders observe how the policy's intended flexibility instead produces systematic ambiguity, shifting the burden of interpretation, technical coordination, and accountability onto stakeholders themselves.

We contribute to CSCW by presenting an illustrative case of how stakeholders across the food supply chain experience the nation-scale implementation of data tracing rules in food production. This case adds to the CSCW literature an example of how the creation of inequitable data work, according to those with less power in the agri-food system, stems from well-intentioned federal policy. More specifically, for agricultural and policy-oriented CSCW scholarship, our study showcases how the imposition of nation-scale, data-driven tracing rules reconfigures diverse agri-food supply chain stakeholders into data workers who must perform invisible labor to sustain the system; identifies three specific points of tension that arise from such reconfigurations; and discusses possible directions through which CSCW could contribute to more equitable data-driven governance practices.

\section{Background}
\subsection{Food Traceability Rule}
\subsubsection{Motivation and Vision}

The US Congress enacted the Food Safety Modernization Act (FSMA) in 2011 in response to “dramatic changes (...) in our understanding of foodborne illness and its consequences, including the realization that preventable foodborne illness is both a significant public health problem and a threat to the economic well-being of the food system” \cite{webpage}. The Act tasked the FDA with nine rules to enhance food safety by establishing shared responsibilities across the supply chain.

Among these rules, the Food Traceability Rule requires supply chain participants of certain fresh foods to maintain detailed tracking records by January 2026 \cite{webinar}. Here, the FDA envisions this data as a “building block” for interoperable technologies, enabling faster identification and removal of contaminated food. It aims to create a harmonized, standardized language for data tracing and management, forming the backbone of FSMA's “new era of tech-enabled traceability” \cite{webinar}. A pilot study with the food industry demonstrated the effectiveness of lot-code-based tracking and removal \cite{mcentire2013pilot,fda2016report}. The Rule was finalized on November 15, 2022, with compliance required by January 20, 2026.

The Rule allows full or partial exemptions, including for small farms with under \$25,000 in annual sales, farms selling directly to consumers, entities holding food for personal consumption, and handlers of foods expected to undergo processing that removes them from “fresh” status. Stakeholders were encouraged to consult the official Rule webpage to confirm exemption status by the FDA \cite{webpage}.

\subsubsection{Implementation of the Food Traceability Rule}

\begin{table}[h]
\centering
\sffamily
\small
% \resizebox{\textwidth}{!}{%

\begin{tabular}{p{9cm}|p{3cm}}
\toprule
         Cheese (made from pasteurized milk), fresh soft or soft unripened& Tomatoes (fresh)\\
 \rowcollight Cheese (made from pasteurized milk), soft ripened or semi-soft&Tropical tree fruits (fresh)\\
 Cheese (made from unpasteurized milk), other than hard cheese&Fruits (fresh-cut)\\
 \rowcollight        Shell eggs& Peppers (fresh)\\
         Finfish (fresh, frozen, and previously frozen), histamine-producing species& Nut butters\\
\rowcollight         Finfish (fresh, frozen, and previously frozen), species potentially contaminated with ciguatoxin& Cucumbers (fresh)\\
         Finfish (fresh, frozen, and previously frozen), species not associated with histamine or ciguatoxin& Herbs (fresh)\\
\rowcollight         Smoked finfish (refrigerated, frozen, and previously frozen)& Leafy greens (fresh)\\
         Crustaceans (fresh, frozen, and previously frozen)& Leafy greens (fresh cut)\\
\rowcollight         Molluscan shellfish, bivalves (fresh, frozen, and previously frozen)& Melons (fresh)\\
         Vegetables other than leafy greens (fresh-cut)& Sprouts (fresh)\\
\rowcollight         Ready-to-eat deli salads (refrigerated)& \\
\bottomrule
    \end{tabular}
    \caption{Food Traceability List (FTL) (2020) \cite{webpage}}
    \label{tab:ftl}
\end{table}

\begin{table}[h]
\centering
\sffamily
\small
% \resizebox{\textwidth}{!}{%

    \begin{tabular}{p{12cm}}
\toprule
                Traceability lot code (TLC) for the food \\
\rowcollight    Quantity and unit of measure of the food \\
                Product description for the food \\
\rowcollight    Location description for the immediate subsequent recipient (other than a transporter) of the food \\
Location description for the location from which you shipped the food \\
\rowcollight    Date you shipped the food \\
Location description for the traceability lot code source or the traceability lot code source reference \\
\rowcollight    Reference document type and reference document number (maintain only) \\
\bottomrule

    \end{tabular}
    \caption{Example KDEs for the shipping CTE}
    \label{tab:kde_example}
\end{table}

The Food Traceability Rule requires stakeholders who manufacture, process, pack, or hold foods on the \textbf{Food Traceability List (FTL)} to maintain traceability records. The FDA developed the FTL with expert panels, the CDC, and an advisory group to identify high-risk foods based on outbreak frequency, illness severity, contamination likelihood, and public health and economic impacts \cite{ftlmodel2022}.

For each initial production of a food item in the FTL (e.g., receiving fish from a vessel or first-time packing of produce), stakeholders must assign a unique \textbf{Traceability Lot Code (TLC)}. They may create these codes autonomously, continuing existing systems if they wish \cite{webinar}. While TLCs need not be kept electronically, in cases of contamination, the FDA will require digital submission via electronic spreadsheets within 24 hours.

Whenever a \textbf{Critical Tracking Event (CTE)} occurs, responsible entities must record the \textbf{Key Data Elements (KDE)} for each TLC. The FDA defines both CTEs and their required KDEs. CTEs include harvesting, cooling, initial packing, first land-based receiving, shipping, receiving, and transformation. The required KDEs differ by the CTEs. For instance, example KDEs for shipping CTE are shown in Table \ref{tab:kde_example}. Through these implementation measures, the FDA aims to establish standardized nationwide traceability data \cite{webinar}.

\subsection{Data Work and Power in CSCW}

In this section, we review how data work has been discussed within the CSCW literature. In particular, we attend to studies that show how data work is often realized through invisible forms of labor, disproportionately burdening those with less power in order to fulfill the visions behind imposed data systems, including in agri-food production contexts.

\subsubsection{Core Concepts of Data Work}

\citeauthor{bossen2019data} define data work as “any human activity related to creating, collecting, managing, curating, analyzing, interpreting, and communicating data” \cite{bossen2019data}. Within this broad scope, CSCW scholarship has emphasized how data work often requires collaboration among diverse stakeholders and intersects with core CSCW concepts. These include data's role as a \textit{boundary object} that facilitates collaboration within data works \cite{lutters2007beyond, star1989institutional}; its reliance on \textit{invisible work} from certain stakeholders to sustain systems \cite{star1999layers, suchman1995making}; and its dependence on \textit{articulation work}, where tasks of tracking, maintaining, and using data must be coordinated across multiple actors \cite{schmidt1992taking, strauss1985work, strauss1988articulation}. Building on these contexts, \citeauthor{lu2021data} identify three strands of CSCW research on data work: (1) its impact on collaborative practices across contexts, (2) its effects on people's work experiences, and (3) its role in shaping—and being shaped by—power dynamics among stakeholders engaged in data practices \cite{lu2021data}. With these foci, CSCW and HCI studies of data work have examined diverse contexts including healthcare \cite{sun2023data,kaziunas2017caring}, education \cite{lu2021data}, remote work \cite{sum2025s}, non-profit organizations \cite{bopp2017disempowered}, and civic engagement \cite{meng2019collaborative}.

In discussing the data work imposed by the FDA's Food Traceability Rule and its impact on stakeholders, we draw particularly on the second and third strands: how visions of “data-driven” efficacy often rely on invisible work, and how this work is frequently imposed on stakeholders with less power.

\subsubsection{Data Work: Laden with Invisible Labor and Power Imbalance}

\citeauthor{muller2019data} note that conversations about “data-driven” systems often focus on abstract processes and pipelines, rendering the actual work and the situated realities of those who perform it less visible \cite{muller2019data}. Building on this, scholars have emphasized the labor-driven nature of data-driven practices and investigated the specific practices that enable such systems \cite{pine2022investigating, sun2023data}. These studies highlight how transforming existing work or living environments into data work environments often requires not only new resources and tools but also the conversion of previously non-data workers into data workers, necessitating new skills and roles \cite{pine2022investigating}. Within such contexts, the work required to sustain data-driven systems, including data collection, analysis, use, and the educational efforts needed to acquire these skills, becomes “invisible work” carried out by individuals and groups \cite{passi2017data}.

When the transformation toward data-driven systems is voluntary, it poses fewer concerns. However, CSCW scholars have highlighted that such implementations often occur within particular power dynamics, disproportionately affecting those who perform invisible work. Beyond being assigned labor that was previously not their responsibility, these power imbalances manifest in forms such as data being used for surveillance (e.g., corporations monitoring employees' remote work practices) \cite{sum2025s}, and participants being compelled to align their activities with data work practices, leading to an “erosion of autonomy” in how they conduct their work \cite{bopp2017disempowered}.

The tension between data-driven visions and the lived realities of those who must perform data work to realize them often becomes more pronounced when datafication is imposed at scale. A prominent example is the imposition of data-driven governance under the guise of societal benefit. Recent studies on COVID-19 contact tracing, for example, have sparked widespread discussion about the ethical implications of large-scale data practices \cite{hall2022supporting, haring2023less, arzt2021tracing, jamieson2022unpacking}. Data-driven governance tools in agriculture have also been critiqued for oversimplifying complex practices and obscuring labor, ultimately reinforcing existing power structures \cite{ghosh2024data, liu2021legibility}. 

In response to the injustices introduced by data-driven governance, \citeauthor{lee2024reconfiguring} emphasize the importance of attending to data relations—the negotiations, tensions, and power dynamics among stakeholders within institutional data infrastructures—as a way to better account for those marginalized by such practices \cite{lee2024reconfiguring}. Together, this body of work illustrates how datafication of everyday work can create new burdens due to mismatches between its intended vision and real-world use, and how CSCW is well positioned to surface these tensions and guide more equitable alternatives. 

\subsubsection{Data Work in Agri-Food Contexts}

The datafication of everyday work in agricultural and food production contexts has been discussed extensively in the field of \textit{precision farming} under \textit{industry 4.0} (or, the ``Fourth Industrial Revolution'') where digital technology, analytics, and cyber-physical systems shape industrial output \cite{Gilchrist_2016}. Precision farming standardizes practices, using site-specific sensing, sampling, and management to treat fields as heterogeneous, cutting costs and conserving resources \cite{Finger_Swinton_Benni_Walter_2019}. More broadly, incorporating ``big data'' into farming alters both stakeholder dynamics and farmers' roles \cite{Wiseman_Sanderson_Zhang_Jakku_2019}.

As a blueprint for datafication efforts, the impact of precision agriculture on farmers has been studied mainly outside the US. \citeauthor{Goller_Caruso_Harteis_2021}'s study of German dairy farmers showed that shifting daily tasks from manual labor to monitoring and controlling technology produced mixed feelings \cite{Goller_Caruso_Harteis_2021}. Farmers appreciated reduced physical work but felt overwhelmed by untrained technology use and maintenance, and reported being perpetually ``on duty,'' causing psychological stress. \citeauthor{Doidge_Palczynski_Zhou_Bearth_van_Schaik_Kaler_2023}'s investigation of UK cattle farmers further revealed a generational ``data divide'' \cite{Doidge_Palczynski_Zhou_Bearth_van_Schaik_Kaler_2023}. The authors describe the ``destabilizing'' effect of data practices, from technological disorientation to increased office work despite farmers' preference for outdoor tasks. Many also struggled with a lack of interlocutors, such as veterinarians, to help interpret data, limiting its value. Critically, \citeauthor{Doidge_Palczynski_Zhou_Bearth_van_Schaik_Kaler_2023} identify three obstacles: farmers unequipped for data work, misaligned temporal and spatial rhythms, and minimal feedback leading to fragmented reproduction practices \cite{Doidge_Palczynski_Zhou_Bearth_van_Schaik_Kaler_2023}.

Additional concerns include farmer privacy under constant monitoring \cite{Kaur_Hazrati_Fard_Amiri-Zarandi_Dara_2022} and a shift in cognition as farmers are increasingly positioned as \textit{cyborgs} \cite{Velden_Klerkx_Dessein_Debruyne_2024}. Some fear standardized agri-environmental systems will increase bureaucratic oversight and benefit others over farmers \cite{Forney_Epiney_2022, Zhang_Heath_McRobert_Llewellyn_Sanderson_Wiseman_Rainbow_2021}. For smallholders, data management is especially challenging; case studies in Tanzania highlight implementation difficulties \cite{Mushi_and_Burgi_2023}. In China, datafication and quantification of farming roles are increasingly reshaping farmer identity \cite{Xie_and_Ma_2019}.

Building on these rich works, our study further demonstrates the specific \textit{tensions} that arise when a national policy imposes data work across US agri-food supply chain stakeholders. Prior work has already identified a range of tensions associated with large-scale data work in food production contexts, including fears of surveillance \cite{sum2025s, Kaur_Hazrati_Fard_Amiri-Zarandi_Dara_2022, Forney_Epiney_2022, Zhang_Heath_McRobert_Llewellyn_Sanderson_Wiseman_Rainbow_2021}, the educational and financial demands of compliance \cite{pine2022investigating, passi2017data}, feelings of being constantly “on duty” \cite{Goller_Caruso_Harteis_2021}, and the erosion of autonomy and identity \cite{bopp2017disempowered, Doidge_Palczynski_Zhou_Bearth_van_Schaik_Kaler_2023}. Our analysis resonates with and adds to this body of work by highlighting moments when compliance with imposed data work becomes infeasible or outright impossible, and situations in which the supposed flexibility of data work produces additional burdens through ambiguity.

\section{Methods} 
To conduct our study, we conducted a qualitative document analysis \cite{altheide2008emergent, bowen2009document} of public comments submitted between 2014 and 2021 to the official docket for the Food Traceability Rule, using Data Feminism \cite{datafeminism} as a theoretical guide. In this section, we first introduce Data Feminism as our methodological background. Then, we introduce the public comments submitted by stakeholders as our data, along with what public comments are, how they were submitted, how we gathered them, and why this corpus is such an important site to attend to under our methodological background. Finally, we present how we analyzed the comment data.

\subsection{Methodological Background: Data Feminism}

Data feminist scholars explore the dynamics between those represented by data and those who collect, analyze, and act upon it. They argue that traditional scientific approaches to data often reproduce existing hierarchies, and call for more intersectional, responsible, and justice-oriented data practices \cite{datafeminism, tran2024situating}. Data feminism has also been applied to understand the transformation of non–data professionals into data professionals. For example, \citeauthor{darian2023enacting} draws on Data Feminism to investigate how nonprofits can resist the three dominant “S's” of datafication—surveillance, selling, and science—in order to remain focused on their missions of social good \cite{darian2023enacting}. 

As such, we adopt Data Feminism as a theoretical orientation to attend to the power (im)balances embedded in the Food Traceability Rule's data tracking and tracing requirements, and to examine how these imbalances manifest as tensions experienced by those required to comply. Our approach is grounded particularly in Principle 1 of Data Feminism: Examine Power \cite{datafeminism}. This principle emphasizes that data is not neutral, but rather representative of dominant power structures and urges researchers to critically examine \textit{who} designs data systems, \textit{who} is excluded, and \textit{who} bears the burden of compliance. 

That said, Data Feminism (1) informed our choice of public comments as a data corpus that reveals the perspectives of those with less power in shaping policy, and (2) oriented our discussions by helping us ask the questions of “Who is doing the work of data science (and who is not)? Whose goals are prioritized (and whose are not)? And who benefits from data science (and who is either overlooked or actively harmed)?” \cite{datafeminism}.

\subsection{Treating Public Comments as a Data Corpus}

Under the Administrative Procedure Act of 1946, federal rulemaking in the United States invites public comments through an open forum \cite{kerwin2018rulemaking}. Today, this process largely takes place on the digital platform Regulations.gov, where the government provides details on proposed and finalized regulations, along with a space for the public to engage by submitting \textit{comments}.

For the Food Traceability Rule, the FDA held two major public comment rounds: the first in early 2014, which focused on defining the scope of the FTL, and the second in September 2020, following the announcement of the Proposed Traceability Rule. The 2020 round included a 120-day comment period and three public meetings. Across both rounds, the FDA received 1,198 comments from a wide range of individuals and entities, including farmers, fishers, restaurant owners, professors, consumers, and blockchain professionals.

In our study, we treat these comments not as “scientifically representative” data of how the Rule affects all stakeholders, but as active grounds through which power dynamics of data work are articulated, resisted, and negotiated \cite{liu2021legibility}, aligning with our Data Feminist orientation. One might raise concerns that public comments over-represent those negatively affected by the Rule, even if the number of beneficiaries is larger. However, as feminist scholars, we invite a shift in analytic focus away from the scale of benefit distribution and toward questions of whose voices are heard or erased, and what those in less powerful positions are articulating. This orientation allows us to more holistically attend to the issue and move toward a more equitable understanding. In other words, rather than viewing public comments as representative of all supply chain actors, we interpret them as situated expressions of critique coming from the margins, or a form of \textit{complaint}: a genre of participation that surfaces tensions for those lacking formal power in regulatory processes, exposing gaps between rule and reality, system and situatedness, pipeline and practice \cite{ahmed2021complaint}.

At the same time, by the same logic, we acknowledge a key limitation of this dataset: these comments likely exclude those who lack the resources, infrastructure, or capacity to engage with the Rule or submit feedback through Regulation.gov, thereby further silencing already marginalized voices. While public comment forums are ostensibly open, participation is neither evenly distributed nor fully accessible. Such an observation raises questions about whether such data captures the experiences of all those burdened by the Rule. For instance, we speculate that Amish communities, who play a significant role in American agriculture but may culturally refrain from engaging with digital technologies, as well as individuals who lack digital literacy or access to devices and may not even be aware that these policies are posted on government websites, are unlikely to have articulated their positions through public comments despite being affected by the Rule. While we observe that some of these positionalities appear indirectly through comments submitted by intermediaries (e.g., retailers who work with Amish communities and articulate concerns on their behalf), we nevertheless acknowledge and articulate the limitations of our comment data.

\subsection{Study Procedure}
\subsubsection{Data Gathering}
The comments on the Rule were submitted in one of three file formats: (1) typed text, (2) attached PDF files, or (3) a combination of both. In some cases, posting entities restricted public access to their comments. From 2014 to 2021, a total of 1,198 comments were submitted by diverse stakeholders, including individuals and groups of food producers, merchants, agricultural experts, and representatives from the technology industry \cite{comments}. All publicly accessible comments were downloaded via the webpage's official bulk-download function, yielding a spreadsheet containing the written comment, the name (s) of the author(s), timestamps, and any attached files. 
 
\begin{figure}
    \centering
    \includegraphics[width=1\linewidth]{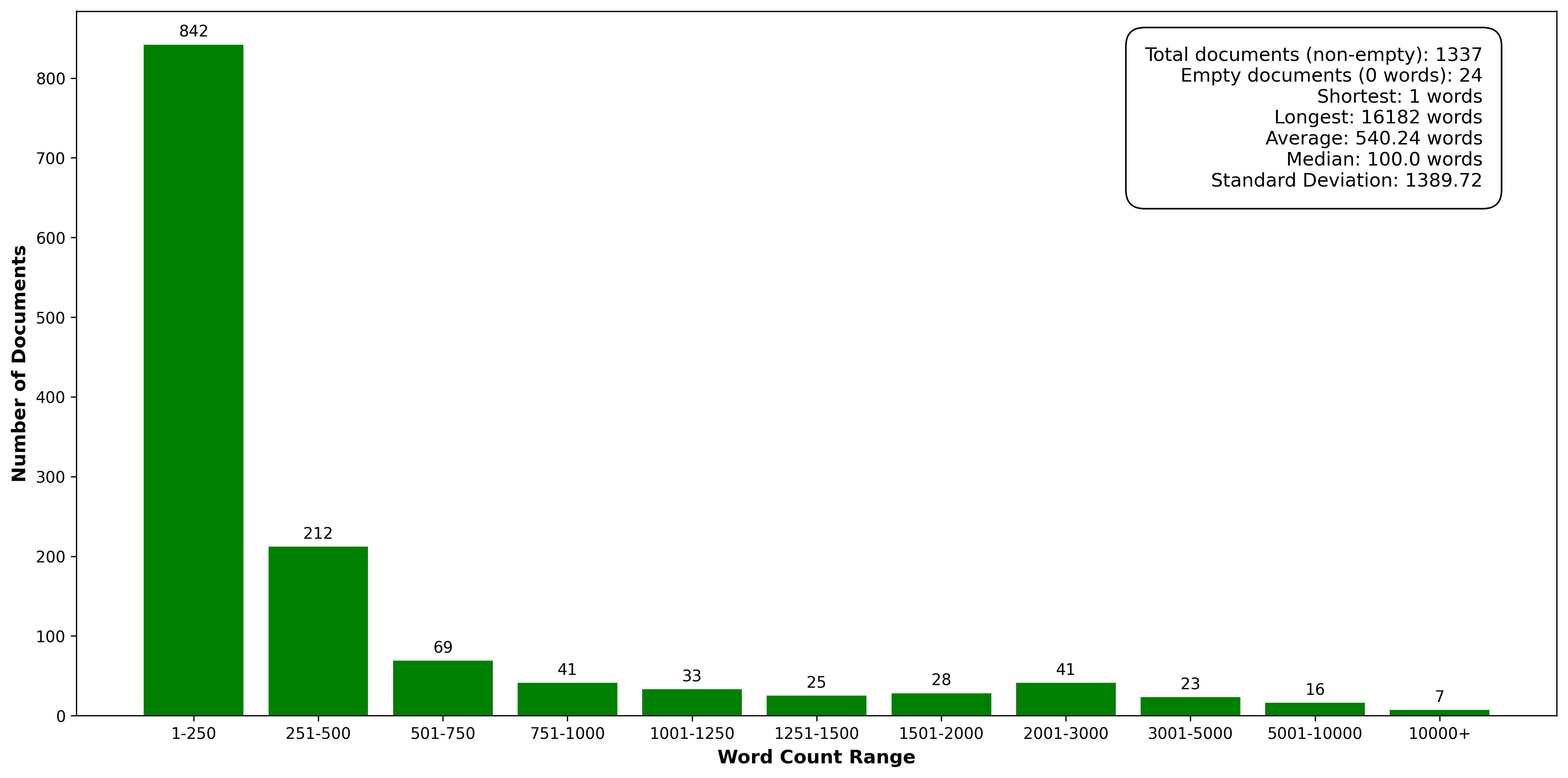}
    \caption{Comment Lengths by Word Count Range}
    \label{Graph}
\end{figure}

As a result, the first author retrieved 1,430 files from the 1,198 total comments. These files varied widely in form and length, ranging from a one-sentence comment containing swear words directed at the government to a 50-page-long comment submitted by a special interest group offering their opinion based on legal and scientific analysis of the Proposed Rule. Of the 361 PDF files, 93 were either broken, duplicates, or could not be transformed to TXT files. As a result, we used the remaining 1,337 files as the main corpus for our analysis. Refer to figure \ref{Graph} for detailed file characteristics.
 
\subsubsection{Analysis}

Using the retrieved comments, we conducted a qualitative document analysis \cite{altheide2008emergent, bowen2009document} to examine how stakeholders responded to the FDA's proposed data-tracing structures. Document analysis involves the systematic review and interpretation of texts to surface patterns, meanings, and institutional logics embedded within written discourse. It is particularly well suited for analyzing bureaucratic records, policy documents, and public engagement with regulatory systems \cite{adie2024digital, carvalho2024central, wong2023privacy}.

Our document analysis consisted of two phases. In the first phase, the first author adopted a bottom-up approach, reading the full corpus of comments and marking (i.e., open-coding) notable excerpts that articulated tensions between the policy's implementation and those required to perform the associated data work. These initial codes took the form of short, descriptive summaries of recurring concerns, such as “collision among existing policies causing confusion about which policy to follow” or “lack of non-technical options to comply with the Rule.” This phase facilitated familiarization with the conversations and recurring issues present across the comment data.

In the second phase, we took a more systematic approach by identifying remarks in which stakeholders discuss specific data work structures being imposed. To do so, we selected five keywords from the Rule that represent core implementation mechanisms of the data work infrastructure mandated by the policy:

\begin{itemize} 
\item \textbf{FTL, Food Traceability List} – A list of foods designated by the FDA for tracking
\item \textbf{TLC, Traceability Lot Code} – Codes assigned by stakeholders to food items on the FTL 
\item \textbf{CTE, Critical Tracking Events} – Supply chain events for which stakeholders are required to document KDEs 
\item \textbf{KDE, Key Data Elements} – Specific data elements recorded for each CTE and linked to a TLC 
\item \textbf{Electronic spreadsheet} – The required format for submitting traceability data in the event of a contamination outbreak 
\end{itemize}

By inputting these keywords—both acronyms (e.g., “FTL,” “CTE,” “KDE”) and their expanded terms (e.g., “critical,” “tracking,” “events”)—into the corpus, we extracted documents in which they appeared and analyzed the surrounding text to examine how commenters interpreted, resisted, or reframed the FDA's proposed system.

\begin{figure} 
\centering 
\includegraphics[width=1\linewidth]{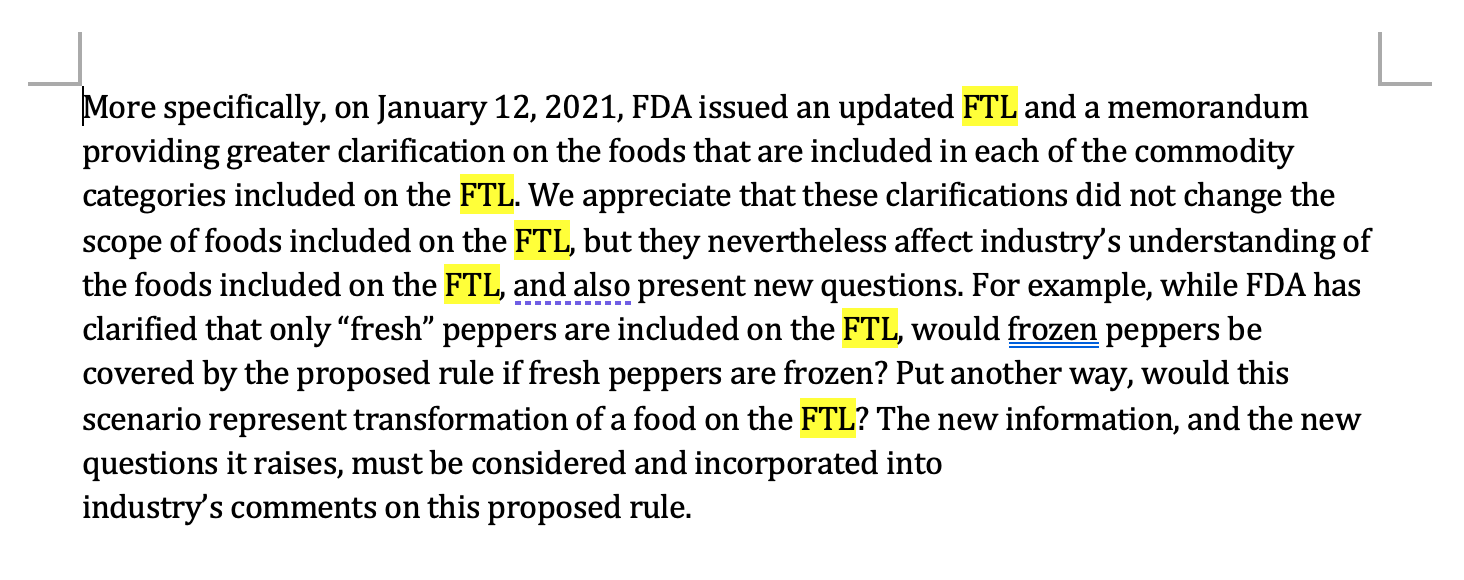} 
\caption{Sample Image of Highlighted Keywords in Comment Files} 
\label{highlight} 
\end{figure}

To support this process, we combined all converted files into a single \texttt{.docx} document, searched and highlighted these keywords, and reviewed the surrounding paragraphs (see Figure \ref{highlight}). While this approach yielded some false positives (e.g., matches such as “affeCTEd” or “seleCTEd”), these were readily identified and dismissed during close reading. Our goal was not to establish exhaustive lexical precision, but rather to enable contextual interpretation through qualitative reading. We treated missed instances due to typos or linguistic variation as an inevitable limitation of document analysis. We mitigated this by incorporating multiple keyword forms and reading broadly around each identified match.

As we examined the text surrounding these keywords, we once again coded notable excerpts, this time with particular attention to how stakeholders reacted to the specific data-work structures imposed by the policy. The first author then refined these codes by integrating them with those developed in the first phase, collapsing overlapping codes and adding those that were overlooked in the keyword-driven analysis. This step was taken to ensure that the second phase, which foregrounded policy jargon, did not obscure or exclude the perspectives of commenters who were less familiar with regulatory terminology.

Finally, the first and second authors collaboratively analyzed the refined codes, clustering related codes into broader themes, identifying patterns across the data, and reflexively interpreting these themes. Through this process, we developed a cohesive account of how the Rule's imposition of data work creates tensions for stakeholders in complying with it.

In analyzing the comments, we did not categorize them by date of submission or submitting entities for several reasons. As noted above, the comments were collected through two primary rounds of public comment: the first in 2014, which focused on the list of foods to be included under the FTL, and the second in 2021, which addressed the Rule as a whole. While we were attentive to the seven-year gap between these two rounds, the overarching policy vision, nor the proposed implementation practices, changed substantially during this period. We also did not observe meaningful differences in commenters' reactions across the two timeframes, including in the specific technologies referenced (e.g., electronic spreadsheets or Microsoft Excel). One exception was a small number of blockchain companies proposing their products as implementation solutions; however, these comments fall outside the scope of our research, which centers on understanding the lived realities of stakeholders currently participating in the food production supply chain. Accordingly, we did not identify substantive differences in comment content across the two periods and therefore did not categorize the comments by temporality.

Furthermore, although the comments were submitted by a wide range of stakeholders, including agri-food supply chain actors, customers, scholars, and technology companies working in traceability, we did not categorize the documents into stakeholder groups prior to analysis. This decision was motivated by two considerations. First, stakeholder identities were often explicitly stated or became apparent through a close reading of the comments. Second, perspectives from non–supply chain stakeholders provided important contextual grounding for interpreting the data holistically, and we did not want to rule them out simply because they fell outside the scope of the agri-food supply chain stakeholders we wanted to focus on.

\section{Findings} \label{finding}
Across the data, we observe a consistent throughline: the Rule does not merely introduce new reporting requirements—it reconfigures food producers and stakeholders as data workers, responsible for generating, managing, and transmitting granular traceability data. This shift, while potentially a necessary step toward public health and interoperability, produces wide-ranging tensions that reflect deep mismatches between policy expectations and on-the-ground realities. 

We organize our findings across three interrelated levels: (1) On an individual level, we showcase how stakeholders experience this role shift, especially through the educational, labor, and financial burdens imposed by data tracking requirement; (2) On the production environment level, we look into how the Rule's implementation collides with the physical, infrastructural, and cultural environments in which food is produced, often rendering compliance infeasible; (3) Finally, on the policy level, we demonstrate how the policy's intended flexibility manifests as ambiguity, placing the burden of interpretation and interoperability onto stakeholders themselves.

\subsection{Tensions in Individual Realities: When Following the Policy Creates Burdens}

In this section, we examine how the Rule is experienced at the individual level by stakeholders, with special attention to the specific \emph{burdens} of complying with the policy. We identify several forms of burden: (1) educational: making sense of the newly introduced Rule and terminology as well as learning to use digital technologies needed for data tracking and keeping, (2) labor: the manual labor of either tracking data or digitizing paper records, and (3) financial: the cost of coping with the previous two forms of burden, along with data storage and management.

Firstly, stakeholders noted that the policy's systemic shift requires substantial time and effort to understand and implement. Importantly, we see comments specific to smaller operations (such as small family farms, small food businesses, and retailers, among others), where stakeholders often lack access to the training, funding, or external expertise available to larger stakeholders (such as large corporate farms), placing them at a disadvantage.

\begin{quote}
``FDA also projects that firms will only need to conduct `an average of 2 hours of training with respect to an average of 3 records.' Again, this is a vast understatement. The proposed rule establishes an entirely new set of key terms and definitions, many of which deviate significantly from current, widespread understandings of similar terms and concepts.'' (Comment from International Foodservice Distributors Association)
\end{quote}

These burdens were felt most acutely during the shift from paper-based to digital recordkeeping, especially challenging for older farmers and those with limited computer literacy who “may not be comfortable with or have the software needed for making electronic spreadsheets.” (Comment from Missouri Coalition for the Environment)

We further note how these educational burdens stem from the reconfiguration of food production experts into data workers. Stakeholders underscore that food producers' expertise lies in agriculture—not in tasks such as extensive data collection, digitization, and management.

\begin{quote}
``[W]e are primarily experts in growing the food, not in collecting extensive sets of data such as GPS coordinates of crop sites [...] and compilation of these data into electronic spreadsheets.'' \textcolor{blue}{(Comment from an Individual Commenter)}
\end{quote}

Furthermore, stakeholders articulated labor (and thus, financial) burden that would follow with the newly imposed data work. This burden was especially articulated by small stakeholders, particularly those just above the \$25,000 exemption threshold—who barely missed the threshold of being classified as small practice but often still lack the resources to manage such tasks. As one commenter explained,

\begin{quote}
``It will affect my VERY SMALL BAKERY. Only 1 person [...] There is no way I can keep track of where the eggs come from or use spreadsheets.'' \textcolor{blue}{(Comment from an Individual Commenter)}
\end{quote}

In addition, many commenters predicted additional labor in converting paper-based records to electronic form when working with entities that are not subject to the Rule. For instance, when downstream entities interact with upstream entities that are exempt from the Rule but maintain paper records, the burden of digitization would fall to the downstream entities. As illustrated by one commenter, 

\begin{quote}
   ``(...) many bakers receive records in non-electronic form, particularly from small suppliers. It would be extremely resource-intensive to assemble a spreadsheet with those records within 24 hours, and this requirement could ultimately force [us] to avoid working with such suppliers.'' \edit{(Comment from American Bakers Association)}
\end{quote}

Commenters also highlighted the financial burden, not only in coping with the aforementioned educational and labor burden through hiring additional staff or specialists, but also in data storage and management. As one producer noted:

\begin{quote}
``The organizing and storing of spreadsheets and data of the magnitude proposed also adds to the financial strain of our small business, as the cost of data storage and data management will exponentially increase.'' \edit{(Comment from Our Harvest Cooperative)}
\end{quote}

Taken together, these concerns illustrate how the Rule's reconfiguration of stakeholders into data workers imposes layered educational, labor, and financial burdens. Rather than a simple shift in recordkeeping practices, it represents a profound restructuring of producers' daily work, perhaps disproportionately affecting those with fewer resources to adapt.

\subsection{Tensions in Production Environments: When Following the Policy is Infeasible or Impossible}

In this section, we showcase how stakeholders' production environments make compliance with the Rule infeasible or impossible. Specifically, the comments report moments where adhering to tech-enabled food traceability is not feasible due to: (1) the (lack of) technological infrastructures, (2) cultural and relational infrastructures that shape ways of being, and (3) the messy and fluid production practices. By doing so, we examine the tensions that arise when the Rule superimposes an idealized version of food production (such as CTEs or KDEs) that does not realistically map onto the varied sites of food production.

\subsubsection{Lack of Technological and Physical Infrastructures}

Many comments noted that the Rule assumes all stakeholders will readily have access to the necessary technological infrastructure to comply. Following long-standing traditions in HCI/CSCW scholarship, we define infrastructures as systems that sustain everyday life, such as power grids that provide households with reliable electricity \cite{star1999ethnography}. Here, we highlight how commenters describe a variety of infrastructural struggles that range from small-scale devices such as laptops to large-scale infrastructures such as Internet connectivity: 

\begin{quote}
“[M]any farmers in rural Missouri are still struggling with reliable Internet - a resource that would be necessary to respond to such a request.” \edit{(Comment from Missouri Coalition for the Environment)}
\end{quote}

Failing to account for the absence of, or difficulty in accessing, technological infrastructure presents cases where it is infeasible or difficult for rural stakeholders to comply with the Rule by positioning urban forms of food production as the norm and marginalizing all others \cite{hardy2019rural}. For example, in the event of food contamination, if the FDA requests a spreadsheet of electronic data within 24 hours, those without Internet access would need to travel to a location with Internet access.

\subsubsection{Cultural Infrastructures Against Tech-Usage}

In addition to physical infrastructure, commenters described sociocultural factors that shape their relationships with technology. In some cases, stakeholders deliberately opt out of digital systems—not because of limited access, but as an expression of religious, environmental, or personal values. As \citeauthor{star1999ethnography} writes, “Infrastructure is both relational and ecological—it means different things to different groups, and it is part of the balance of action, tools, and the built environment, inseparable from them” \cite[p.473]{star1999ethnography}. Here, the infrastructure of food production is shaped as much by identity and ways of being as by material conditions. This perspective aligns with HCI and CSCW literature on non-use \cite{baumer2015importance}, which frames disengagement not as a failure, but as a valid and intentional mode of participation.

Perhaps the most prominent example would be the Amish\footnote{The Amish community is a religious group known for deliberately prohibiting or strictly limiting digital technologies, a stance that HCI and CSCW scholarship has frequently examined as an important case of intentional non-use.} stakeholders. The American Cheese Society comments that “there are cheesemakers among our membership that are Plain Sect (e.g., Amish), and do not use electronic recordkeeping or any other electrical devices for their business,” and asks the FDA to include alternate means of acceptable recordkeeping for these processors.

Such intentional disengagement from technology, or non-use, is also showcased outside of religious contexts. For one, it may appear simply as a choice of volitional technical (non-)use \cite{rothschild2024swapping}, deeply resonating with previous scholarship, where farmers felt the erosion of their identities when being reconfigured into data workers \cite{Doidge_Palczynski_Zhou_Bearth_van_Schaik_Kaler_2023}. As one commenter noted: “These are farmers, not computer techs. This is why some get into farming—to leave the wilds of technology behind, to some regard.” \textcolor{blue}{(Comment from an Individual Commenter)}

In some cases, stakeholders' desire to step away from technology was driven by the tension between the implicit obligations required to comply with the Rule and their identities. For instance, Ozark Forest Mushrooms, LLC commented on how sustainable farms may not be able to comply with the packaging practices required to be considered exempt from the Rule, as “some commonly used containers that are recyclable are not allowed to be used,” and the FDA's recommendation “that to meet this exemption, farms individually wrap each cucumber in sealed plastic wrap with the appropriate labeling” would produce immense environmental waste, which goes against their identity as a sustainable farm.

These moments illustrate that the feasibility of data tracking extends beyond physical infrastructure; it also involves cultural ways of being that are deeply tied to identity and must be recognized and respected, especially when new practices are imposed. Echoing the previous section's critique of defaulting to certain technological environments, we highlight how assuming technological comfort as the norm creates a gap between the envisioned system of data tracing and the lived realities of those tasked with implementing it.

\subsubsection{Production Practice}

Another form of tension between the FDA's idealized data work environment and the practical realities of stakeholders' work arose from the way operationalized data work units—such as clear-cut CTEs or KDEs—failed to account for the plural and fluid realities of food production. Commenters emphasized that food production rarely conforms to neatly defined processes. In this section, we illustrate several of these moments among the many examples raised by commenters.

Commenters noted that there are production practices that do not fit clear-cut CTEs, creating situations where data tracking becomes infeasible or confusing. Cross-docking serves as a notable example. Cross-docking is the practice in which products pass through a facility without being formally accepted into inventory or managed as shipments or receipts. As commenters noted: 

\begin{quote} 
“Cross docking is another circumstance where the `shipper' and `receive' CTEs do not appropriately fit the circumstances. Cross dock vendors never accept product into their inventories, and their records of transactions with transporters are typically paper based. As a result, cross dock vendors do not have the capabilities to serve as a shipper or receiver.” \edit{(Comment from FMI- The Food Industry Association)} 
\end{quote}

As an activity that falls into the ambiguous space between shipping and receiving—where food items are handled but not processed in ways that could introduce contamination—it remains unclear whether cross-dockers are subject to the rule, and whether they are required to record KDEs for receiving, shipping, or both.

Commenters also emphasized how the “ever-changing dynamics of supply chains” complicate the determination of whether an entity is acting as a first receiver or a receiver, creating additional burdens in determining and collecting the appropriate KDEs for their situation. \edit{The Food Industry Association notes:}

\begin{quote} 
``Many entities cannot apply the Proposed Rule's distinction between the CTEs of 'receiving' and ‘first receiving' due to the day-to-day variations in product sourcing. When a retailer places an order for produce through a broker, for example, the produce may arrive at the retailer's location through multiple different avenues, including directly from the farm or from a warehouse. Even when a retailer orders the same product from the same broker, the supply chain may vary from day to day or week to week.'' \edit{(Comment from FMI- The Food Industry Association)}
\end{quote}

Commenters themselves highlighted how these clear-cut units seem to rest upon the FDA's assumption of a particular production practice (large industrialized farms rather than smaller ones), showing how the imposed data work practices unevenly impact stakeholders depending on their positionalities. One such example concerns the requirement to record the physical location of production in geographic coordinates. As the Farm and Ranch Freedom Alliance noted:

\begin{quote} 
``The requirement to keep records that include the `growing area coordinates' are based on assumptions about large-scale, largely monoculture agricultural operations, in which a single crop is grown on multiple acres. Diversified sustainable farms operate very differently. Crops are often interspersed with each other, and their planting locations rotated frequently. Depending on the climate, multiple plantings of the same crop may occur in the same year, but in slightly different areas of the farm.”  (Comment from Farm and Ranch Freedom Alliance)
\end{quote}

Failing to account for these practices that fall outside of clear-cut units resting on the FDA's assumed and idealized data work environment can impose undue burdens: requiring duplicate data work where intra-organizational systems already exist but are not recognized by the Rule, or mandating data collection in cases where the risk of contamination is minimal. In these scenarios, the Rule risks creating unnecessary compliance data work without corresponding public health benefits.

\subsection{Tensions in Policy Design: When Promised Flexibility Turns Into Ambiguity}

In this section, we explore how the \emph{flexibility} intended by the Rule inadvertently becomes a form of \emph{policy ambiguity}, producing confusion, infrastructural misalignments, and new forms of labor for stakeholders. The only explicit technological requirement under the Rule is the use of an electronic spreadsheet to report traceability data within 24 hours of a contamination event. To maintain "flexibility," the FDA does not mandate the use of specific software or methods for recordkeeping—perhaps in keeping with the broader FSMA promise to avoid prescribing specific technologies.

\begin{figure}
    \centering
    \includegraphics[width=1\linewidth]{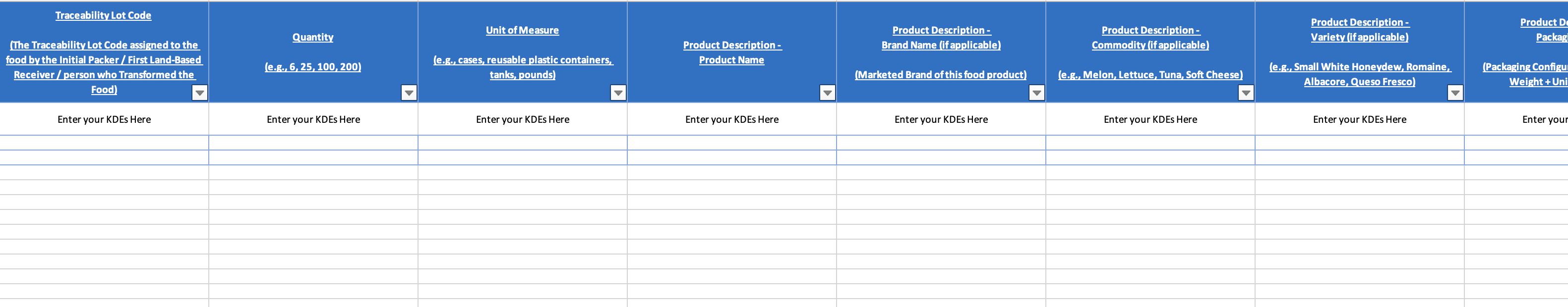}
    \caption{Sample Excel Spreadsheet Template Provided by the FDA}
    \label{fig:excel}
\end{figure}

However, as commenters made clear, this flexibility introduces uncertainty and shifts the burden of interpretation, design, and implementation onto stakeholders. Although the FDA provides a sample Excel spreadsheet template, its use is not required, and the file itself is simply a blank sheet with labeled columns for KDEs and CTEs—far from a functional standard (see fig. \ref{fig:excel}). Without clear guidance, stakeholders are left to guess how to operationalize it, inputting data in whatever format they choose. As one commenter points out,

\begin{quote} 
"There will be a need to hire people to create and maintain a database system for electronic recordkeeping, even if it can be an Excel spreadsheet that is searchable by FDA upon request. This is because it is not clear just what is needed for the spreadsheet." (Comment from Alaska Seafood Marketing Institute)
\end{quote}

To address these concerns, commenters consistently requested greater clarity on expectations for electronic recordkeeping. Several suggested that the FDA provide standardized templates to guide compliance. They discussed how the lack of clear direction regarding technology use risks placing additional labor and financial burdens on stakeholders. Without knowing precisely what is required, they may need to hire personnel to develop and maintain database systems, even for basic compliance. Stakeholders further articulated how a lack of clarity may risk wasting resources on systems that may ultimately be incompatible with their supply chain partners. National Association of Convenience Stores \& Society of Independent Gasoline Marketers of America notes:

\begin{quote} 
``The lack of any technological standards for the entire supply chain strongly disincentivizes firms from participating in the Agency's traceability efforts, because the Proposal would force firms to invest in traceability software, equipment, and training in the two years before the regulations go into effect without knowing whether their systems will be interoperable with the traceability systems of other firms.'' \edit{(Comment from National Association of Convenience Stores \& Society of Independent Gasoline Marketers of America)}
\end{quote}

Ironically, this lack of coordination undermines the very vision the FDA promotes: a \emph{tech-enabled, interoperable traceability system}. For instance, if stakeholders turn to commercial off-the-shelf solutions to avoid compliance failure, this may unintentionally result in \emph{informal vendor lock-in}, centralizing compliance pathways around a few dominant platforms and contradicting the Rule's spirit of decentralization and innovation.

More critically, the Rule's ambiguity appears to obstruct the creation of the “harmonized data language” it seeks to establish. Several commenters emphasized that interoperability depends not only on shared incentives but on shared standards—something the Rule defers: 

\begin{quote} 
"FDA's Proposed Rule also neglects to address how the Traceability Lot Codes that are to contain the Key Data Elements from Critical Tracking Events will be shared by firms across the food supply system, otherwise known as interoperability. [...] Establishing interoperability within the system needs to be done in tandem with setting data standards, not afterwards." (Comment from National Association of Convenience Stores \& Society of Independent Gasoline Marketers of America)
\end{quote}

Of course, some stakeholders welcomed the flexibility embedded in the Rule, resonating with broader concerns discussed earlier about avoiding the generalization of a singular, idealized data-tracking experience. Several commenters appreciated the Rule's accommodation of diverse business practices:

\begin{quote} 
``The Proposed Traceability Rule remains flexible for those entities and the farms that sell to them and makes certain the creation of a lot code is not cost prohibitive. [...] All these options ensure scale-appropriate flexibility for small businesses and short supply chains. [We] request that FDA keep this flexibility around what KDEs can contain, including the definitions of `physical location name' and `location identifier.' '' \edit{(Comment from National Sustainable Agriculture Coalition)}
 \end{quote}

However, flexibility without structure is not the same as inclusivity. Across many comments, a clear concern emerged: the absence of shared standards leads to duplication of effort, misaligned investments, and additional labor. These burdens fall disproportionately on those least equipped to absorb them, ultimately undermining both the technical feasibility of a seamless traceability pipeline and the public health outcomes it is meant to generate. In sum, we find that what is framed as flexibility becomes a systemic ambiguity, a policy condition that externalizes the cost of technical coordination to individual stakeholders. 

This aligns with broader critiques that highlight how policy vagueness can serve both political and administrative purposes \cite{abebe2020roles, kaplow2013rules}. When policymakers are not fully informed about the conditions on the ground, they may opt for broadly applicable rules that, while non-optimal for any single context, aim to achieve aggregate efficiency across diverse situations \cite{mahoney2005general}. In such cases, the responsibility for ensuring that policies are implemented in ways that reflect stakeholders' lived realities falls to administrative agencies—such as the FDA—which are tasked with applying that flexibility with care \cite{eskridge2014cases}.

However, even if the FDA seeks to implement such tailored guidance, doing so may be constrained by FSMA's broader framework, which explicitly prohibits mandating specific technologies. Directing stakeholders to use particular systems could contradict the intent of the statute. This results in a contradictory situation: a Rule designed to offer inclusivity and flexibility instead generates ambiguity—and, ultimately, additional burdens for those it was meant to support. In this contradiction, we witness how well-intentioned flexibility—without adequate scaffolding—can transform from a gesture of inclusivity into a mechanism of exclusion, burdening the very stakeholders it was meant to empower.

\section{Discussions}
Our findings examined how the FDA's Food Traceability Rule reconfigures agri-food supply chain stakeholders into data workers. In doing so, we highlighted three tensions that arise from such reconfigurations: (1) the educational, labor, and financial burdens that come with becoming data workers; (2) the challenges of complying with the Rule due to production realities (including physical, infrastructural, and cultural limitations); and (3) the confusion and burden stakeholders experience due to the ambiguity of the Rule.

In our Discussion, we revisit our findings through a Data Feminist lens to highlight how these tensions are experienced disproportionately by different stakeholders and to argue that those reconfigured to data workers should be recognized not as collateral damage in the pursuit of data-driven visions, but as essential policy partners. We then draw on our team's interdisciplinary background in HCI, law, and economics to examine why addressing these tensions is, in many respects, structurally difficult within existing policy implementation frameworks. Finally, we outline ways the CSCW community might engage with these tensions to support more democratic and equitable forms of data-driven policy work.

\subsection{Data Feminist Perspectives on Stakeholders in Data-Driven Policy: Recognizing and Respecting Them as \textit{Policy Partners}}

Our findings focused on how the reconfiguration of stakeholders into data workers produces bodily, cognitive, logistical, and financial burdens, echoing longstanding CSCW scholarship on the invisible labor required to sustain sociotechnical systems \cite{Bowker_Star_2008}. A key discussion point we look into is that such burdens are usually most acutely experienced by smaller stakeholders; in other words, the Rule disproportionately affects its subjects, resonating with prior works on how data-driven governance often oversimplifies on-the-ground practices and reinforces existing power imbalances, while neglecting marginalized realities \cite{lee2024reconfiguring, liu2021legibility}.

In this section, drawing on Data Feminism, we examine how data work, such as those imposed by the Food Traceability Rule, can rarely be implemented in a truly “equal” manner. Data Feminist scholarship asks: “Who is doing the work of data science (and who is not)? Whose goals are prioritized (and whose are not)? And who benefits from data science (and who is either overlooked or actively harmed)?” \cite[p. 47]{datafeminism}. These questions provide a lens for critically examining who benefits from particular data practices and who is left behind, directing attention to aspects of data systems that demand greater care to achieve more equitable data practices. Accordingly, we ask who is overlooked in the FDA's Food Traceability Rule, in service of which goals, and how these oversights might be mitigated by positioning stakeholders as policy partners rather than as actors simply subject to compliance.

\subsubsection{Who is overlooked in the Food Traceability Rule?}
In complying with the data work imposed, some stakeholders were inevitably positioned more advantageously than others, with these gaps likely to widen over time. As \citeauthor{datafeminism} note, “\emph{Equality is measured from a starting point in the present… But this formula for equal treatment means that those who are ahead in the present can go further, achieve more, and stay on top, whereas those who start out behind can never catch up}” \cite[p.31]{datafeminism}. This insight echoes work in infrastructural HCI and critical data studies that cautions against one-size-fits-all systems and highlights how standardized infrastructures can obscure difference while reproducing inequality \cite{Bowker_Star_2008, datafeminism, ghosh2024data}. 

This dynamic is clearly reflected in our findings, particularly in two ways: (1) how categorization within datafication is disproportionately experienced by different stakeholders, and (2) how the ability to comply with the Rule is disproportionately shaped by whether the material, infrastructural, and organizational prerequisites for compliance are already in place.

Categorization has long been central to processes of datafication, as it imposes discrete rulings on practices that are often continuous, fluid, and open-ended. As Foucault argues, the creation and enforcement of categories constitutes a form of power \cite{Foucault_1989}. Within CSCW, \citeauthor{Bowker_Star_2008} similarly show that systems of classification are not merely technical mechanisms but social arrangements that shape people, their labor, and their relationships \cite{Bowker_Star_2008}. For example, nursing taxonomies in the United States were developed not to support nurses' everyday work but to facilitate documentation and the tracking of activities \cite{Henry_Mead_1997, Henry_Holzemer_Randel_Hsieh_Miler_1997}. While such systems enabled more precise billing and performance analytics for administrators and institutions, they failed to account for the realities of nurses' work, including forms of labor that fell outside prescribed categories as well as the additional labor required to produce these data.

Our findings show similar dynamics taking place for the stakeholders under the Food Traceability Rule. We observed multiple cases where stakeholders located near categorical margins were particularly disadvantaged: Small practices just above the \$25,000 exemption threshold were required to comply with the Rule despite lacking the resources typically associated with larger operations; The Rule's Critical Tracking Events (CTEs) required stakeholders to fit messy, fluid, and non-linear production practices into discrete categories such as “freezing” or “shipping,” rendering forms of production that did not neatly align with these categories effectively invisible; Although not explored in depth in our findings, many stakeholders also raised concerns about how fresh foods were defined for inclusion in the Food Traceability List (FTL), debating ambiguous cases such as nut butters and other borderline products.

Taken together, these cases illustrate how the Rule transforms a broad spectrum of production scales, practices, and food items into segmented categories such as CTEs, FTLs, and KDEs. While these categories enable standardization and enforcement needed for datafication, they also generate ambiguity or even miscategorization for those positioned at their boundaries. For stakeholders operating at the margins of these classifications, these ambiguities and miscategorizations are far more likely to translate into compliance risks, additional labor, and material loss.

Our findings further show that some stakeholders are far better positioned to absorb and adapt to the changes required by the Rule than others. Large stakeholders with the financial capacity to absorb new costs, industrialized operations with clear-cut and standardized workflows, actors located in regions with well-established infrastructure, and those already performing data work are likely able to comply with the Rule with comparatively less tension. In other words, what appears as a neutral compliance requirement is, in practice, deeply political: while more resourced actors can accommodate this shift, smaller stakeholders experience the resulting financial and operational burdens disproportionately.

These experiences echo scholarship in rural computing and infrastructural HCI, which has long shown how infrastructural gaps produce unequal access to and participation in sociotechnical systems \cite{palacios2024mending, jonas2022digital}. Viewed through a Data Feminist lens, our findings show that the Rule privileges stakeholders with the resources to comply with newly imposed data work, while those with limited infrastructure, staffing, or purchasing power are forced to stretch their capacities under conditions of uncertainty. In short, what appears as equal treatment at the level of policy produces inequitable outcomes in practice.

\subsubsection{Not collateral damage, but integral policy partners}

As \citeauthor{datafeminism} emphasize in Principle 1 of Data Feminism, “\emph{The matrix of domination works to uphold the undue privilege of dominant groups while unfairly oppressing minoritized groups}” \cite[p. 26]{datafeminism}. Combined with our reflection on how small-sized stakeholders are systematically positioned to fall behind the Rule's requirements, we must now critically ask what these costs are ultimately for.

The data work imposed by the Food Traceability Rule is largely justified in the name of public health. However, the costs and responsibilities of realizing this data-driven vision rest on the invisible labor performed by stakeholders, imposing disproportionate tensions and burdens on small-sized stakeholders in particular. We argue that these stakeholders do not bear primary responsibility for public health; rather, public health is a governmental goal and responsibility. Stakeholders instead function as collaborators upon whom the state relies to realize this data-driven vision of public health. In this sense, they must be understood as \emph{policy partners} whose cooperation is requested, rather than as actors simply subjected to compliance obligations.

The implementation of data-driven policy visions inevitably depends on someone's invisible labor \cite{muller2019data}. However, treating such labor as an unfortunate but necessary byproduct risks obscuring both the existence and value of those who perform this work, as well as the conditions under which it is carried out. Instead, these forms of labor must be made visible and recognized, and the voices of those who perform them must be more actively heard in order to mitigate the tensions and burdens they experience as they are reconfigured into data workers. This includes creating mechanisms where stakeholders can articulate their burdens and receive meaningful support as crucial actors needed to realize the data-driven vision, rather than being positioned as passive recipients of regulatory demands.

Yet the Rule not only positions stakeholders as actors expected to comply with its requirements (rather than as crucial partners in realizing its vision) but also offers few avenues for stakeholders to voice their lived realities or access meaningful and visible support. Beyond subsidization for compliance, the primary support mechanisms available to stakeholders include the ``Small Entity Compliance Guide'', a 38-page, jargon-heavy document buried within the FDA's website that itself imposes significant cognitive and learning burdens \cite{FDA2023TraceabilityGuide}, and the FSMA Technical Assistance Network (TAN). Inquiries submitted through TAN are handled on a case-by-case basis, with unclear timelines and limited personalization, offering little practical relief for stakeholders navigating compliance in real time.

In light of stakeholders' reconfiguration into data workers, we argue that this shift must be explicitly recognized—not as collateral damage, but as evidence that these actors are integral partners in realizing policy goals. Treating them as such requires making their burdens visible, listening to their experiences, and actively developing forms of support that respect their labor and capacities, rather than assuming compliance as a given.

\subsection{Evaluating the Tensions: Legal and Administrative Constraints}
In this section, we draw on our team's interdisciplinary perspectives—HCI, CSCW, law, and economics—to explore the broader issue of envisioning and implementing data-driven governance policies. In doing so, we consider how laws, policymakers, those who implement policies, and those who must comply often face different and sometimes conflicting concerns and constraints, directing our attention to the realistic challenges that arise in these discussions. We first highlight tensions among three forces: Congress's legislative vision, federal administrative implementation, and stakeholders' everyday experiences. We then examine the potential, limits, and implications of public comment gathering in mitigating these tensions, framing it as a digital meeting ground for collaborative navigation.

One of the biggest obstacles to flexible, robust data pipelines is often the law itself. In our findings, small, rural, and culturally distinct stakeholders noted that lacking clear and shared standards under the notion of ``flexibility'' creates duplication, misaligned investments, and extra labor. Yet legislators and regulators often view flexibility as a feature. When policymakers lack on-the-ground knowledge, they favor broad rules that, while suboptimal locally, aim for aggregate efficiency \cite{mahoney2005general}. In such cases, responsibility shifts to agencies like the FDA to apply flexibility with care, ensuring policies reflect lived realities \cite{eskridge2014cases}. This reflects critiques that policy vagueness functions as both a political and administrative tool \cite{abebe2020roles, kaplow2013rules}.

The federal agency seeks to address this vagueness by using its internal data and expertise to propose rules that implement Congress's vision. However, in our case, even tailored guidance is constrained by FSMA, which states that the “requirement shall not prescribe specific technologies for the maintenance of records” \cite{FSMA2011}. Directing stakeholders to use particular systems could therefore contradict the statute's intent or congressional mandate. This creates a contradiction: a rule designed for inclusivity and flexibility instead generates ambiguity and added burdens. In this way, well-intentioned flexibility, without adequate scaffolding, can shift from a gesture of inclusivity to a mechanism of exclusion, burdening the very stakeholders it sought to empower.

Beyond statutory limits, the agency sought public comment to gather diverse perspectives and ensure participatory, transparent, and effective regulations. The process, however, does little to mitigate existing power dynamics. As noted, in formulating the Traceability Rule, the FDA assumed digital capacity, interoperability, and data formats despite many comments to the contrary, effectively centering large, tech-enabled operations as the norm. Under the Administrative Procedure Act, agencies must consider all substantive arguments from commenters \cite{USC553c2018}. Yet policymakers retain broad discretion, allowing them to ignore complaints from smaller or differently organized stakeholders even when those contest agency assumptions.

Whether agencies consider commenter \textit{preferences}, as distinct from substantive arguments, is often left to agency discretion. Most agencies give little weight to comments based on preferences, values, or sentiments—a category that includes many of those described above. \citeauthor{farina2012rulemaking} argue that because agency experts pursue rulemaking in a deliberative and technical manner, expressions of preference are often inappropriate, especially when not representative \cite{farina2012rulemaking}. Rulemaking, after all, is not a vote \cite{TipsEffectiveComments}. Yet by privileging only “representative” views, agencies reinforce assumptions of universality and overlook difference—an approach that runs counter to Data Feminism.

Studies of civic tech failures further show that participation should be designed to meet people where they are, not where institutions assume \edit{\cite{hamm2024does,hagan2019participatory,gilman2022beyond}}. Yet current U.S. governance models—legislation, regulation, and participation—were never built for this goal. This dynamic, where legislators, federal agencies, and stakeholders of varying scales, cultures, and beliefs work within the same system toward the shared aim of preventing foodborne illness despite differing constraints, closely echoes CSCW's emphasis on collaborative work enabled by sociotechnical infrastructures \cite{lutters2007beyond, star1989institutional}. From this perspective, the following section considers how CSCW is uniquely positioned to advance discussions on supporting such messy yet vital forms of collaboration, and what our future works might entail.

\subsection{CSCW's Role in Data-Driven Governance}
Building on Sections 5.1 and 5.2, we now turn to the role the CSCW community might play in engaging with these challenges. In Section 5.1, we argued that data-driven governance often inevitably reconfigures certain actors into data workers, and that responding to this reality requires (1) recognizing and respecting these actors and their labor, (2) actively listening to the tensions and burdens they experience, and (3) developing meaningful forms of support to address those tensions and burdens. In Section 5.2, we further examined why governance mechanisms themselves often struggle to fulfill these goals. Against this backdrop, this section explores the three ways that CSCW can contribute to moving forward: advocating for stakeholders' roles as policy partners rather than mere subjects of compliance; supporting the visibility and articulation of the tensions and burdens they experience; and helping design sociotechnical interventions that meaningfully respond to and mitigate these burdens in practice.

\subsubsection{Capturing tensions between tech-driven policy visions, implementation, and situated realities}

First, we discuss how CSCW can play an active role in recognizing and giving voice to the actors who perform data work created by data-driven policies. For policymakers, as well as legal and regulatory scholarship, our recognition of the invisible data work performed by stakeholders carries important implications. Currently, in policymaking, the additional labor entailed by proposed rules is primarily accounted for through assessments of ``paperwork burden''—defined as the time, effort, or resources required to generate, maintain, or disclose information to a federal agency \cite{CareyOrtiz2024}. Yet our work shows that this definition fails to capture the implicit and invisible data labor identified in our findings—the interpretive, technical, and clarification work required to navigate and comply with data- or technology-driven mandates. These findings offer policymakers, as well as legal and regulatory scholars, new lenses for interpreting and assessing policies, contributing to more equitable forms of data-driven governance.

Against this backdrop, we position CSCW as a field uniquely equipped to identify, recognize, and give voice to these erased practices in policymaking. Prior work in policy-oriented HCI and CSCW has repeatedly argued that policy can no longer be treated as separate from technology, and that policies that use or govern technologies should be positioned at the center, rather than the periphery, of CSCW inquiry \cite{yang2023designing, manuel2020place}.

One way of advancing this agenda is by leveraging CSCW's longstanding attention to the situated experiences of those subjected to technical infrastructures—whether material systems or infrastructures instantiated through regulation—and its ability to surface tensions and frictions between a system's intended goals, its implementation, and lived realities. Grounded in people-centered views of technology and policy, as well as its trans-disciplinary capacity to bring diverse fields together to account for technical feasibility, user values, governmental processes, and public interests \cite{gairola2025public,yang2024future}, CSCW can help surface policy tensions and inform more responsive sociotechnical policies \cite{spaa2019understanding}. Our study further demonstrates how longstanding feminist approaches in HCI, CSCW, and data studies—being attentive to erased or invisible realities—are well-suited to capturing the tensions produced by data- and technology-driven policy infrastructures. 

Accordingly, we envision CSCW collaborating more actively with those who shape policy realities by foregrounding its strengths and contributing to more equitable data- and technology-driven policymaking, including through collaboration with policy-shaping actors such as think tanks \cite{spaa2019understanding}.

\subsubsection{\textcolor{blue}{Envisioning public participation systems that allow stakeholder communication and connection}}

Another role that CSCW can play is empowering stakeholders to communicate their tensions and burdens more meaningfully \textcolor{blue}{both to the governmental bodies and among themselves}. In our study, this communication primarily occurred through the public comment system, which served as the core site of analysis. While Regulations.gov is intended as a mechanism for democratic engagement, we find that, in practice, it is often ineffective for \textcolor{blue}{all potentially affected} stakeholders. The platform is heavily online-based, limiting access and legibility for those without reliable digital infrastructure; it accepts input only within a narrow, agency-defined time window; and presents comments in board-like formats that obscure who is speaking, what positions are articulated, and whether collective voices are forming.

CSCW and HCI scholarship have long examined democratic participation in increasingly technologized contexts, including studies of digital platforms that shaped same-sex marriage deliberation in Taiwan \cite{bardzell2020join} and systems designed to support community participation in policy designs \cite{matias2018civilservant, kuo2025policycraft}. Building on this work, we argue that public comment systems have unrealized potential as participatory infrastructures that can better support democratic communication and visibility among stakeholders affected by data-driven policies.

One concrete opportunity lies in enabling connections among commenters. Notably, our dataset included numerous near-identical comments submitted by different individuals (including a case where the phrase “Customize your comment and help FDA understand why this matters:” remained unedited), suggesting moments of collective mobilization that remained largely invisible within the platform interface. Beyond such mobilized efforts, many stakeholders articulate similar concerns, yet the current system prevents them from seeing, recognizing, or building upon one another's voices, as commenters are represented solely through the total number of comments submitted.

Building on these observations, we may envision public participation systems that enable solidarity among similarly positioned stakeholders, not only to amplify collective voices but also to make participation safer by distributing risk across a group rather than isolating individuals \cite{ahmed2021complaint}. Such efforts may further build on prior work showing that democratic participation must incorporate experiential, cultural, and intersubjective forms of expression that foster connection, shared understanding, and the mitigation of coercion \cite{bardzell2020join}, as well as scholarship demonstrating how sociotechnical systems can support collective action in response to data-driven reconfigurations of work \cite{ming2025surveillance}. 

At the same time, such representations must remain attentive to less visible forms of consensus, avoiding the erasure through abstraction \cite{bardzell2020join}. Any reimagining of public participation must attend to stakeholders who are unable or unwilling to engage through digital platforms \textcolor{blue}{and thus would not be served by improvements in visibility and connection within the platform}. Reliance on online systems as the primary mode of participation risks reinforcing existing exclusions. Town halls, mailed submissions, and other non-digital forms of participation may continue to play important roles here. The challenge, then, is to envision participatory systems capable of integrating and communicating across both digital and non-digital modes of engagement.

Building on these lessons, CSCW scholarship can explore how public participation systems might be reimagined—not simply as mechanisms for gathering comments, but as democratic infrastructures that support communication among affected communities and meaningful visibility into how participation shapes policy outcomes.

\subsubsection{Supporting Stakeholders as They “Make Do” With Imposed Data Work}

Another important consideration is the practical reality that the data work imposed by the Food Traceability Rule is unlikely to disappear in the near term. Even when such tensions are somehow acknowledged by policymakers, policy change often unfolds slowly. In this context, alongside ongoing community efforts to advocate for more equitable data-driven policies, imagining how the tensions and burdens produced by imposed data work might be mitigated in the short term constitutes an important and attainable role for CSCW.

CSCW research has long examined how individuals subjected to adverse technical and informational infrastructures find ways to cope with them, demonstrating autonomy, agency, and subjectivity \cite{bardzell2015user,vertesi2014seamful}. Within the domain of data work, scholars have shown how teachers resist and renegotiate autonomy when classroom interactions are datafied, and how remote workers under data-driven management employ diverse tactics—from technological workarounds to peer solidarity—to reclaim agency \cite{sum2025s}. 

In this light, CSCW can direct attention to how stakeholders in the agri-food supply chain might be supported in addressing the three tensions identified in our findings. This may involve designing tools, systems, workflows, or sociotechnical supports that help stakeholders adapt to, negotiate with, collectively clarify, or redistribute the burdens of compliance \cite{ming2025surveillance}. Importantly, such tools need not be designed as one-size-fits-all solutions mandated by government agencies; rather, they allow for greater plurality in design, where different tools are developed for different contexts, and adoption would be voluntary rather than universal or mandated.

At the same time, the practical limits of this direction must be acknowledged. Only a small fraction of systems proposed in academic venues or under the name of ‘civic tech' reach stable, long-term use \cite{hamm2024does}, due to various reasons such as lack of sustained funding. Introducing fee-based models risks imposing new financial burdens on already constrained producers, while alternative paths—such as private funding, partnerships, or open-source development—present their own challenges. Moreover, tools that rely on digital infrastructures may remain inaccessible to stakeholders who lack such resources. These constraints raise important questions about the extent to which CSCW and HCI interventions can meaningfully address structural inequities through design alone, underscoring the need for careful consideration of where such ``solutions'' are feasible and where their limits lie.

\section{Conclusion}
In this study, we examined how the FDA's Food Traceability Rule—while aspiring toward a tech-enabled future of healthier food systems—reconfigures stakeholders into data laborers, often disproportionately burdening those with limited resources to comply. This shift reveals critical gaps between data-driven governance visions and lived realities. We invite the CSCW community to recognize our unique position in surfacing these tensions as interdisciplinary fields, and to actively engage in envisioning and advancing participatory, justice-oriented approaches to tech- and data-driven policy design that center the experiences of those most often overlooked.

\begin{acks}

We thank Cindy Kaiying Lin for her valuable support in writing this work. The work was supported in part by NSF grant 2418059.

\end{acks}

\bibliographystyle{ACM-Reference-Format}
\bibliography{01Reference}

\received{May 13, 2025}
\received[revised]{January 13, 2026}
\received[accepted]{March 17, 2026}

\end{document}
\endinput
%%
%% End of file `sample-authordraft.tex'.